\ifnum\pdfoutput=0\relax
\documentclass[
    a4paper,
    fleqn,
    envcountsect,
    envcountsame,
    dvisvgm,
    hypertex
  ]{llncs}
\else
\documentclass[
    a4paper,
    fleqn,
    envcountsect,
    envcountsame,
  ]{llncs}
\fi

\usepackage{hyperref}
\hypersetup{%
  pdftitle={Offline Drawing of Dynamic Trees: Algorithmics and Document Integration},
  pdfauthor={Malte Skambath and Till Tantau},
}

\usepackage{listings}

\usepackage{
  amssymb,
  amsmath,
  amsfonts,
  wrapfig
}

\usepackage[utf8]{luainputenc}

\usepackage{wrapfig}
\newbox\mybox
\newdimen\mydimen

\usepackage{tikz}

\ifnum\pdfoutput=0\relax%
\else
  \usetikzlibrary{external}
  \tikzset{external/system call={lualatex \tikzexternalcheckshellescape -halt-on-error
  -interaction=batchmode -jobname "\image" "\texsource"}}
\fi

\usepackage{url}
\def\TikZ{Ti\emph kZ}
\def\examplestart{
  \begingroup\par
  \unskip\small\smallskip\noindent
  \begin{minipage}[t]{.68\textwidth}
    \vskip-\medskipamount
}
\def\examplemid{
    \vskip-\medskipamount\smallskip\leavevmode
  \end{minipage}\hskip.5em\color{black!25}\vrule width 0.8pt\color{black}\quad
  \begin{minipage}[t]{.265\textwidth}\raggedright\leavevmode\hrule
    width0pt height0pt
}
\def\exampleend{\vskip.5\smallskipamount\end{minipage}\smallskip\par\endgroup}

\lstdefinestyle{example}{language=tikz,basicstyle=\ttfamily\small, columns=flexible}

\newcommand\fixbox[2]{
  \setbox\mybox=\hbox{#1}
  \mydimen=\ht\mybox
  \advance\mydimen by-#2
  \ht\mybox=\mydimen
  \box\mybox
}

\lstdefinelanguage{tikz}
{
  keywordsprefix=\\,
  morekeywords={},
  sensitive=false,
  morecomment=[l]{\%},
}

\lstdefinelanguage{lua}
{
  showstringspaces=false,
  morekeywords={and,break,do,else,elseif,end,false,for,function,if,in,local,
     nil,not,or,repeat,return,then,true,until,while},
   morekeywords={[2]arg,assert,collectgarbage,dofile,error,_G,getfenv,
     getmetatable,ipairs,load,loadfile,loadstring,next,pairs,pcall,print,
     rawequal,rawget,rawset,select,setfenv,setmetatable,tonumber,tostring,
     type,unpack,_VERSION,xpcall},
   morekeywords={[2]coroutine.create,coroutine.resume,coroutine.running,
     coroutine.status,coroutine.wrap,coroutine.yield},
   morekeywords={[2]module,require,package.cpath,package.load,package.loaded,
     package.loaders,package.loadlib,package.path,package.preload,
     package.seeall},
   morekeywords={[2]string.byte,string.char,string.dump,string.find,
     string.format,string.gmatch,string.gsub,string.len,string.lower,
     string.match,string.rep,string.reverse,string.sub,string.upper},
   morekeywords={[2]table.concat,table.insert,table.maxn,table.remove,
   table.sort},
   morekeywords={[2]math.abs,math.acos,math.asin,math.atan,math.atan2,
     math.ceil,math.cos,math.cosh,math.deg,math.exp,math.floor,math.fmod,
     math.frexp,math.huge,math.ldexp,math.log,math.log10,math.max,math.min,
     math.modf,math.pi,math.pow,math.rad,math.random,math.randomseed,math.sin,
     math.sinh,math.sqrt,math.tan,math.tanh},
   morekeywords={[2]io.close,io.flush,io.input,io.lines,io.open,io.output,
     io.popen,io.read,io.tmpfile,io.type,io.write,file:close,file:flush,
     file:lines,file:read,file:seek,file:setvbuf,file:write},
   morekeywords={[2]os.clock,os.date,os.difftime,os.execute,os.exit,os.getenv,
     os.remove,os.rename,os.setlocale,os.time,os.tmpname},
   sensitive=true,
   morecomment=[l]{--},
   morecomment=[s]{--[[}{]]--},
   morestring=[b]",
   morestring=[d]'
}

\lstdefinelanguage{pseudocode}{
  morekeywords={
    if,then,else,while,do,repeat,until,seq,seqdo,return,call,
    for,pardo,in,foreach,print,output,input,exit,
    break,loop,end,begin,goto,pardo,par,global,read,write,to,stop,idle,procedure,function
  },
  sensitive=true,
  morecomment=[l]{//},
  morestring=[b]",
}

\lstdefinestyle{text}{
  basicstyle=\ttfamily\footnotesize,
}

\lstdefinestyle{pseudocode}{
  language=pseudocode,
  numberstyle=\scriptsize\color{black!50},
  numbers=left,
  basicstyle=\rmfamily,
  keywordstyle=\bfseries\itshape,
  columns=fullflexible,
  mathescape=true,
  identifierstyle=\itshape,
  literate={<}{{$<$}}1
           {>}{{$>$}}1
           {-}{{-}}1
           {+}{{$+$}}1
           {*}{{$\cdot$}}1
           {<-}{{$\gets$\ }}2
           {<=}{{$\leq$}}1
           {>=}{{$\geq$}}1
           {!=}{{$\neq$}}1
           {==}{{$=$}}1
}

\newcommand\Class[1]{\mathchoice{\text{\normalfont\small$\mathrm{#1}$}}{\text{\normalfont\small$\mathrm{#1}$}}{\text{\normalfont$\mathrm{#1}$}}{\text{\normalfont$\mathrm{#1}$}}}       
\newcommand{\Lang}[1]{\ifmmode{\text{\textsc{#1}}}\else\textsc{#1}\fi}

\def\linktosvg#1{%
  \ifnum\pdfoutput=0\relax#1\else%
  \href{\svgurl\#\thepage}{#1}%
  \fi%
}

\newtheorem{algorithm}[theorem]{Algorithm}
\spnewtheorem*{criterion*}{Criterion}{\upshape\bfseries}{\itshape}

{\catcode`\~=11
\gdef\svgurl{http://www.informatik.uni-kiel.de/~msk/pub/2016-dynamic-trees/main.html}%
}

\title{Offline Drawing of Dynamic Trees:
  Algorithmics~and~Document~Integration\thanks{Animations in this
    document will only be rendered in the
    \ifnum\pdfoutput>0 \expandafter\href\expandafter{\expandafter\svgurl\expandafter}\fi{\textsc{svg} version}~\cite{SkambathT2016-svg}, see
    Section~\ref{section-output} for a discussion of the reasons.}}  

\author{Malte Skambath\inst{1} \and Till Tantau\inst{2}}

\institute{%
  Department of Computer Science, Kiel University, Germany\\
  \email{malte.skambath@email.uni-kiel.de}
  \and
  Institute of Theoretical Computer Science, Universit\"at zu L\"ubeck, Germany\\
  \email{tantau@tcs.uni-luebeck.de}
}

\begin{document}

\maketitle

\begin{abstract}
  While the algorithmic drawing of static trees is well-under\-stood and
  well-supported by software tools, creating animations depicting how
  a tree changes over time is currently difficult: software support,
  if available at all, is not integrated into a document production
  workflow and algorithmic approaches only rarely take temporal
  information into consideration. During the production of a
  presentation or a paper, most users will visualize how, say, a
  search tree evolves   over time by manually drawing a sequence of
  trees. We present an   extension of the popular \TeX\ 
  typesetting system that allows users to specify dynamic
  trees inside their documents, together with a new algorithm for
  drawing them. Running \TeX\ on the documents then results in
  documents in the \textsc{svg} format with visually pleasing embedded
  animations. Our algorithm produces animations that
  satisfy a set of natural aesthetic criteria when possible. On the
  negative side, we show 
  that one cannot always satisfy all criteria simultaneously and that
  minimizing their violations is NP-complete.
\end{abstract} 

\section{Introduction}

% The core problem
\emph{Trees} are undoubtedly among the most extensively studied graph
structures in the field of graph drawing; algorithms for drawing trees
date back to the origins of the
field~\cite{Knuth1971,WetherellS1979}. However, the 
extensive, ongoing research on how trees can be drawn efficiently,
succinctly, and pleasingly focuses on either drawing a single,
``static'' tree or on interactive drawings of ``dynamic'' 
trees~\cite{CohenBTT1995,CohenBTTB1992,Moen1990}, which are trees 
that change over time. In contrast, the problem of drawing dynamic trees
\emph{non}interactively in an \emph{offline} fashion has received 
less attention.

It is this problem that lies at the heart of our paper. 

\begin{figure}[htpb]
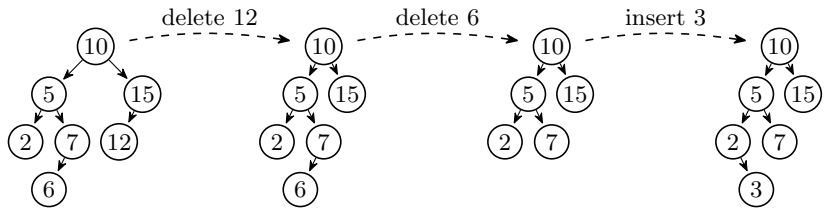

\hfil  \tikz {
  \scoped [graph]
    \graph [binary tree layout, significant sep=0pt] {
      a1/10 -> { 5 -> { 2 , 7 -> 6 } , 15 -> 12}
    };

  \scoped [graph,xshift=3cm]
    \graph [binary tree layout, significant sep=0pt] {
      a2/10 -> { 5 -> { 2 , 7 -> 6} , 15 }
    };

  \scoped [graph,xshift=6cm]
    \graph [binary tree layout, significant sep=0pt] {
      a3/10 -> { 5 -> { 2 , 7} , 15 }
    };

  \scoped [graph,xshift=9cm]
    \graph [binary tree layout, significant sep=0pt] {
      a4/10 -> { 5 -> { 2 -> {,3} , 7} , 15 }
    };

  \draw [shorten <=2mm, shorten >=2mm, >={Stealth[round]}, semithick,
  ->, bend left=10, dashed]
    (a1) edge ["delete 12"] (a2)
    (a2) edge ["delete 6"] (a3)
    (a3) edge ["insert 3"] (a4);  
  }
\caption{A ``manually'' created drawing of a dynamic tree: Each tree
  in the sequence has been drawn using the
  Reingold--Tilford~\cite{ReingoldT1981}  algorithm.}
\label{fig-manual}
\end{figure}

\begin{figure}[htpb]
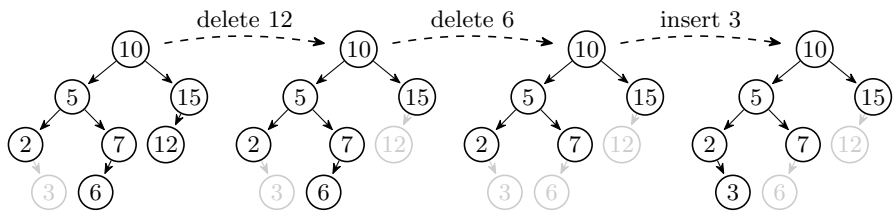

  \hfil
  \tikz [g/.style={draw=black!20,text=black!20}] {
  \scoped [graph]
    \graph [binary tree layout, significant sep=0pt] {
      a1/10 -> { 5 -> { 2 ->[g] {,3[g]}, 7 -> 6 } , 15 -> 12}
    };

  \scoped [graph,xshift=3cm]
    \graph [binary tree layout, significant sep=0pt] {
      a2/10 -> { 5 -> { 2 ->[g] {,3[g]}, 7 -> 6 } , 15 ->[g] 12[g]}
    };

  \scoped [graph,xshift=6cm]
    \graph [binary tree layout, significant sep=0pt] {
      a3/10 -> { 5 -> { 2 ->[g] {,3[g]}, 7 ->[g] 6[g] } , 15 ->[g] 12[g]}
    };

  \scoped [graph,xshift=9cm]
    \graph [binary tree layout, significant sep=0pt] {
      a4/10 -> { 5 -> { 2 -> {,3}, 7 ->[g] 6[g] } , 15 ->[g] 12[g]}
    };

  \draw [ shorten <=2mm, shorten >=2mm, >={Stealth[round]}, semithick,
  ->, bend left=10, dashed]
    (a1) edge ["delete 12"] (a2)
    (a2) edge ["delete 6"] (a3)
    (a3) edge ["insert 3"] (a4);  
  }
  \caption{The dynamic tree from Figure~\ref{fig-manual}, redrawn by
    drawing a ``supergraph'' (the union of all trees in the sequence)
    and then using the positions of nodes in this supergraph for the
    individual drawings.}
  \label{figure-super}
\end{figure}

% An example of what we want to do
Consider how an author could explain, in a paper or in a presentation,
how a tree-based data structure such as a search 
tree works. In order to explain the dynamic behavior, our author
might wish to show how the data structure evolves for a
sequence of update operations. A typical drawing of the evolving
sequence might look as in Figure~\ref{fig-manual}, which has been
created ``manually'' by running the Reingold--Tilford
algorithm~\cite{ReingoldT1981} on each tree in the sequence
independently. While the result is satisfactory, there are (at least)
two shortcomings: 

\emph{First Shortcoming: Flawed Layout.}
In the first step, the layout of the root's children
changes (their horizontal distance decreases) even
though there is no structural change at the root. While in the
present graph the effect is small, one can construct examples where
a single node removal causes a change in distances
on all levels, obscuring where the actual structural change occurred. 
Since the whole sequence of trees (the whole ``dynamic tree'') is
given by the author, the problem can be addressed by not running the
Reingold--Tilford algorithm on each tree individually, but by running
it on the ``supergraph'' resulting from uniting all trees in the
sequence, resulting in the visualization in Figure~\ref{figure-super}.

Unfortunately, this simple supergraph approach introduces new
problems: First, the nodes ``2'' and ``7'' are unnecessarily far apart
-- the nodes ``3'' and ``6'' could use the same space since they are
never both members of the same tree. Second, it is easy to construct
sequences of trees whose union is not a tree itself.

We address these problems in Section~\ref{section-algo}, where we
present a new algorithm for computing layouts of dynamic trees that
addresses the above problems. For dynamic trees whose supergraphs are
trees or at least acyclic, the algorithm finds an optimal layout (with
respect to natural aesthetic criteria) of the dynamic tree in
linear time. For cyclic supergraphs, which are also important in
practice since they arise for instance from the rotations necessary to
balance search trees in data structures such as \textsc{avl}
trees~\cite{AdelsonL1962}, we show that one has to break the cycles in order to
layout the graph according to the criteria we develop. While we show
that it is $\Class{NP}$-complete to find a minimal set of break points, a simple
greedy heuristic for finding breakpoints turns out to produce visually
pleasing results.

\emph{Second Shortcoming: Presentation as a Sequence of Snapshots.}
In order to depict the evolving nature of her dynamic tree, our author
depicted different ``snapshots'' of the tree at different times and
arranged these snapshots in a sequence.  While the temporal dimension
\emph{needs} to be turned into something else when our medium of 
communication is printed paper, for documents presented using
appropriate electronic devices we can visualize dynamic trees using
\emph{animations.} Such an animation needs much less space 
on a page and, perhaps more importantly, our visual system is
\emph{much} better at spotting movement than at identifying structural
changes between adjacent objects. 

In Section~\ref{section-anim} we present a system for creating
animations on-the-fly during a run of 
the \TeX\ program on a text document: First, we have augmented the
popular \TikZ\ graphic package~\cite{Tantau2015} (a macro package for \TeX\ for
creating graphics) by commands that compute and embed animations in
the output files. Due to the way the system 
works, these commands have almost no overhead regarding compilation
speed or resulting file size. Second, we have implemented a prototype
of our algorithm from Section~\ref{section-algo} for drawing dynamic
trees  that uses these animation commands. In result, when an author
specifies the above dynamic graph appropriately in a \TeX\ document
and then runs \TeX\ on it to convert it, the resulting file will
contain the normal text and graphics as well as an embedded animation
of the dynamic tree. When the document is viewed on electronic devices
with a modern browser, the animation runs right inside the
document.
%We detail our arguments in Section~\ref{section-anim} why an 
%integrated approach to creating animations of dynamic trees in
%preferable over specialized software tools and explain 
%how animations are created by our system.

\paragraph*{Related Work.}

Approaches to drawing \emph{static} trees date back to the early
1970s, namely to the work of Knuth, Wetherell and 
Shannon, and Sweet \cite{Knuth1971,WetherellS1979,Sweet1977}. A standard
algorithm still in use today is due to Reingold and
Tilford~\cite{ReingoldT1981}, see also~\cite{Walker1990}. They
suggested that symmetric tree structures should be drawn symmetrically
and provided an algorithm that supports this objective well and runs
in linear time. Instead of visualizing trees as node-link diagrams,
one can also use tree maps~\cite{JohnsonS1991}, three
dimensional cone trees~\cite{RobertsonMC1991}, or
sunburst visualizations~\cite{StaskoZ2000}.

Approaches to drawing general \emph{dynamic} graphs are more
recent. The se\-quen\-ce-of-snapshot visualizations sketched before as
well as animations are standard ways of visualizing
them~\cite{FedericoAMWZ2011}. One can also generally treat  
time as another spacial dimension, which turns nodes into tubes
through space~\cite{GrohHW2009}. There are many further 
techniques that are not restricted to node-link diagrams
\cite{BurchBW2012,BurchD2008,GreilichBD2009,RedaTJLB2011}; for an
extensive overview of the whole state of the art including a 
taxonomy of different visualization techniques see Beck
et~al.~\cite{BeckBDW2014}, or \cite{GrahamK2010} for a more tree-specific overview.  
Diehl, Görg and Kerren~\cite{DiehlGK2000,DiehlGK2001} introduced a
general concept, called \emph{foresighted layout}, for drawing dynamic
graphs offline. They propose to collapse nodes in the supergraph that
never exist at the same time and to then draw the supergraph. While
this approach produces poor results for trees, the results are better
for hierarchical graphs \cite{GorgBPD2004}. 

Approaches tailored specifically to drawing dynamic \emph{trees} are
currently almost always \emph{online} approaches. The algorithms,
which expect a sequence of update operation as
input~\cite{Moen1990,CohenBTTB1992}, are 
integrated into interactive software and create or adjust the layout 
for each change. An early algorithm designed for dynamic trees was
developed by Moen~\cite{Moen1990}. Later Cohen et al.\
\cite{CohenBTT1995,CohenBTTB1992} presented algorithms for different
families of graphs  the includes trees. 
% Heine et.\ al used the supergraph method to draw trees with acyclic supergraph, which 
% obscure the nature of trees by not garuanteeing planar drawings.
% \cite{HeineSFS2006}

Concerning the integration of tree drawing algorithms into text
processing software, first implementations for the typesetting system
\TeX{} date back to Eppstein~\cite{Eppstein1985} and  Brüggemann and
Wood~\cite{BrueggemannKleinW1989}. A more recent implementation of the
Reingold--Tilford algorithm by the second author is now part of the
graph drawing engine in \TikZ{}~\cite{Tantau2013}.

\paragraph*{Organisation of this Paper.}

This paper is structured into two parts: In the first part,
Section~\ref{section-anim}, we present the system we have developed
for generating animations of dynamic graphs that are embedded into
documents. Our core argument is that the system's seamless integration into
a widely used system such as \TeX\ is 
crucial for its applicability in practice. In the second part,
Section~\ref{section-algo}, partly as a case study, partly as a
study of independent interest, we investigate how a dynamic tree can
be drawn using animations. We derive aesthetic criteria that
animations and even image sequences of dynamic trees \emph{should}
meet and present an algorithm that \emph{does} meet them. Full proofs
can be found in the appendix, which also contains a gallery of
dynamic trees drawn using our prototype.

\section{Dynamic Trees in Documents}
\label{section-anim}

The problem for which we wish to develop a practical solution in the
rest of this paper is the following: Visualize one or more dynamic
trees inside a document created by an author from some manuscript. To
make the terminology precise, by \emph{dynamic graph} we refer to a sequence
$(G_1, \dots, G_n)$, where each $G_i = (V_i, E_i, \phi_i)$ is a
directed, annotated graph with vertex set~$V_i$, edge set~$E_i$, and an
annotation function $\phi_i \colon 
V_i \cup E_i \to A$ that assigns additional information to each node
and edge from some set~$A$ of annotations like ordering or size
information. A \emph{dynamic tree} is a dynamic graph 
where each $T_i$ is a tree with the edges pointing away from the
root. A \emph{manuscript} is a plain text written by an \emph{author}
that can be transformed by a program into an (output) \emph{document,} 
a typically multi-page text document with embedded graphics
or embedded animations. Note that the problem is an \emph{offline}
problem since the manuscript contains a full description 
of the dynamic graph and algorithms have full access to it. 
In rest of this section we explain how the practical obstacles arising
from the problem are solved by the system we have developed, in
Section~\ref{section-algo} we investigate algorithmic questions.  

In the introduction we saw an example of how a dynamic tree can be
visualized using a series of ``snapshots'' shown in a row. While this
way of depicting a dynamic tree is a sensible, traditional way of
solving the problem (drawings on printed paper ``cannot
change over time''), documents are now commonly also read on electronic
devices that are capable of displaying changing content and, in
particular, animations. 
We claim that using an animation instead of a sequence of snapshots
has two major advantages: First, sequences of snapshots need a lot of
space on a page even for medium-sized examples. We did a cursory survey
of standard textbooks on computer science and found that typically
only three to four snapshots are shown and that the individual trees
are often rather small. For an animation, the length of the sequence
is only limited by the (presumed) attention span of the reader and not
by page size. Second, our visual system is \emph{much} better at spotting
movement than at identifying structural changes between adjacent
objects. When operations on trees such as adding or deleting a leaf or
moving whole subtrees are visualized using movements, readers can
identify and focus on these operations on a subconscious level. 

Given the advantages offered by animations, it
is surprisingly difficult to integrate animations into documents. Of
course, there is a lot of specialized software for creating animations
and graphics output formats like \textsc{pdf} or \textsc{svg} allow
the inclusion of movie files in documents. However, this requires
authors to use -- apart from their main text processor like \TeX\
or Word -- one or more programs for generating animations and they
then have to somehow ``link'' the (often very large) outputs of
these different programs together. The resulting workflows are
typically so complicated that authors rarely employ them. Even
when they are willing to use and integrate multiple tools into
their workflow, authors face the problem that using different tools makes
it next to impossible to keep a visually consistent appearance of the
document~\cite{Tantau2013}. Very few, if any, animation 
software will be able to render for instance \TeX\ formulas inside
to-be-animated nodes correctly and take the sizes of these formulas
into account.

We have developed a system that addresses the above problems; more
precisely, we have augmented an existing system that is in wide-spread 
use -- \TeX\ -- by facilities for specifying dynamic trees, for
computing layouts for them, and for generating animations that are
embedded into the output files. Our extensions are build on top of \TikZ's
graph drawing engine~\cite{Tantau2013}, which has been part
of standard \TeX\ distributions since 2014.

Authors first specify the dynamic trees they wish to
draw inside \TeX\ manu\-scripts using a special syntax, which we
describe in Section~\ref{section-input} (conceptually,
this is similar to  specifying for instance formulas inside the \TeX\
manuscripts). Next, authors apply a graph drawing algorithm to the
specified dynamic graph by adding an appropriate option to the
specification and then running the \TeX\ program as explained in
Section~\ref{section-processing}. Lastly, in
Section~\ref{section-output}, we discuss which 
output formats are supported by our system, how the output can be
viewed on electronic devices, and how a fallback for printed paper can 
be generated.

\subsection{The Input: Specifying Dynamic Trees}
\label{section-input}

In order to make dynamic trees accessible to graph drawing algorithms,
we first have to specify them. For dynamic graphs and, in particular,
for dynamic trees, there are basically two different methods available
to us: We can specify each graph or tree in the dynamic graph
sequence explicitly. Alternatively, we can specify a sequence of
update operations that transform one graph into the next such as, for
the dynamic trees of search trees, the sequence of insert and delete
operations that give rise to the individual trees. Besides being
easy and natural to use, the second method also provides algorithms
with rich semantic information concerning the change from 
one graph to the next in the sequence.

Despite the fact that the second method is more natural in several
contexts and more semantically rich, for our prototype we use the
first method: Authors specify dynamic graphs by explicitly specifying
the sequence of graphs that make up the dynamic graph. We have two
reasons for this choice: First, specifying the sequence of graphs
explicitly imposes the least restrictions on what kind of dynamic
graphs can be drawn, in principle. In contrast, the set of update
operations necessary to describe the changes occurring just for the
standard data structures balanced search trees, heaps, and
union--find trees is large and hard to standardize. For
instance, should the root rotation occurring in \textsc{avl} trees be
considered a standard update operation or not? Second, it easy to
convert a sequence of update operations into a 
sequence of graphs, while the reverse direction is harder and,
sometimes, not possible. Our system can easily be extended to accept
different sequences of update operations as input and convert them
on-the-fly into a sequence of graphs that is 
then processed further. 

There are different possible formats for specifying individual
graphs and, in particular, trees of graph sequences, including
\textsc{graph\-ml}, an \textsc{xml}-based markup language; the
\textsc{dot} format, used  by \textsc{graphviz}~\cite{EllsonGKNW2004};
the \textsc{gml} format, used by the Open Graph Drawing
Framework~\cite{ChimaniGJKMS2007}; or the format of the
\lstinline[style=example]|\graph| command of \TikZ~\cite{Tantau2015},
which is similar to the \textsc{dot} format. As 
argued in~\cite{Tantau2013}, it is not purely a matter of taste, which
format is used; rather, good formats make it easy for humans to notate
all information about a graph that is available to them. For instance,
for static graphs the \emph{order} in which vertices are specified is
almost never random, but reflects information about them that the
author had and that algorithms should take into account.

Since the algorithm and system we have implemented are build on top of the
graph drawing engine of \TikZ~\cite{Tantau2013}, we can use all of the
different syntax flavors offered by this system, but authors will
typically use the \lstinline[style=example]|\graph| command. Each 
graph in the sequence of graphs is surrounded by curly braces and,
following the opening brace, we say
\lstinline[style=example]|[when=|$i$\lstinline[style=example]|]| to indicate that we now specify  
the $i$th graph in the sequence. The graph is then specified by
listing the edges, please see \cite{Tantau2013} and~\cite{Tantau2015}
for details on the syntax and its use in \TikZ. The
result is a specification of the dynamic graph such as the following
for the example graph from Figures~\ref{fig-manual}
and~\ref{figure-super}: 

\begin{lstlisting}[style=example]
\tikz \graph {   {[when=1] 10->{ 5->{ 2,              7->6 },   15->12 } },
                 {[when=2] 10->{ 5->{ 2,              7->6 },   15 } },
                 {[when=3] 10->{ 5->{ 2,              7 },      15 } },
                 {[when=4] 10->{ 5->{ 2->{ , 3 },     7 },      15 } } }; 
\end{lstlisting}

\subsection{Document Processing and Algorithm Invocation}
\label{section-processing}

Once a dynamic graph has been specified as part of a larger \TeX\
document, we need to process it. This involves both running a dynamic
graph drawing algorithm to determine the positions of the nodes and the
routing of the edges as well as producing commands that create the
desired animation.

The framework provided by the graph drawing engine~\cite{Tantau2013} of
\TikZ\ is well-suited for the first task. All the author has to do is
to load an appropriate graph drawing library and then use a special
key with the \lstinline[style=example]|\tikz|
command:

\examplestart
\begin{lstlisting}[style=example]
\tikz [animated binary tree layout]
  \graph {  {[when=1] 10->{ ... } };
            {[when=2] 10->{ ... } },
            {[when=3] 10->{ ... } },
            {[when=4] 10->{ ... } } };
\end{lstlisting}
\examplemid
\linktosvg
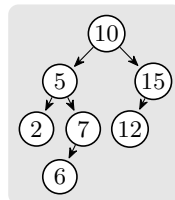
{
\tikz [graph, bg=10, make snapshot if necessary=0] \graph [animated binary tree layout, nodes={fill=white}] {
  {[when=0s] 10->{5->{2, 7->6}, 15->12}},
  {[when=2s] 10->{5->{2, 7->6}, 15}},
  {[when=4s] 10->{5->{2, 7},    15}},
  {[when=6s] 10->{5->{2->{,3}, 7}, 15}}}; 
}
\exampleend
\smallskip

The key \lstinline[style=example]|animated binary tree layout| causes
the graph drawing engine to process the dynamic 
graph. It will parse the dynamic graph, convert it to an 
object-oriented model, and pass it to an algorithm from the
\lstinline[style=example]|evolving| library, which is
written in the Lua programming
language~\cite{Ierusalimschy2006}.\footnote{When the algorithm is
  also implemented in the Lua language, it can be used directly by \TeX\
  without special configurations or runtime linking, but it can also be
  implemented in C or C++ at the cost of a more complicated
  deployment.} The  
framework also handles the later rendering of the nodes and edges and
their correct scaling and embedding into the output
document. Thus, the algorithm's implementation only needs to
address the problem of computing node positions from an
object-oriented model of the dynamic graph. The implementation need not
(indeed, cannot) produce or process graphical output and
primitives.

Once the algorithm has computed the positions for nodes and edges of
the graphs in the sequence, actual animations need to be generated. For this,
\TikZ\ itself was extended by a new animation subsystem, which
can be used independently of the graph drawing engine and allows users
to specify and embed arbitrary animations in their
documents. The animation subsystem adds \emph{animation
  annotations} to the output file, which are statements like ``move
this graphics group by 1cm to the right within 2s'' or ``change the
opacity of this node from opaque to transparent within 200ms.''
More formally, they are \textsc{xml} elements in the Synchronized
Multimedia Integration Language~\cite{Bultermanetal2008}. For the
animation of dynamic graphs, the graph drawing engine can now map the
computed positions of the nodes at different times to \TikZ\ commands
that add appropriate movement and opacity-change annotations 
to the output.

\subsection{The Output: Scalable Vector Graphics}
\label{section-output}

The annotation-based way of producing animations has two important
consequences: Firstly, adding the annotations to the output does not
have a noticeable effect on the speed of compilation (computing the 
necessary \textsc{xml} statements is quite easy) nor on the file size
(annotations are small). However, secondly, the job of rendering the
graph animations with, say, 30 frames per second does not lie with
\TeX, but with the viewer application and we need both a format and
viewer applications that support this. 

Currently, there is only one graphics format that supports these
annotation-based animations: The \emph{Scalable Vector Graphics}
(\textsc{svg}) format~\cite{Dahlstroemetal2011}, which is a general
purpose graphics language that is in wide-spread use. All modern 
browsers support it, including the parsing and rendering of
\textsc{svg} animations. The \textsc{dvisvgm} program, which is part of standard
\TeX\ distributions, transforms arbitrary \TeX\ documents
into \textsc{svg} files that, when viewed in a browser, are visually
indistinguishable from \textsc{pdf} files produced by \TeX\ -- except,
of course, for the animations of the dynamic graphs. 

\label{section-snapshots}%
While we argued that animations are a superior way of visualizing
dynamic graphs, there are 
situations where they are not feasible: First, documents \emph{are}
still often printed on paper. Second, the popular \textsc{pdf} format
does not support annotation-based animations and, thus, is not able to
display \TikZ's animations. Third, it is sometimes desirable or necessary to
display ``stills'' or ``snapshots'' of animations at interesting time
steps alongside the animation.  
 In these situations, authors can say
 \lstinline[style=example]|make snapshot of=|$t$ to
 replace the animation by a static picture of what the animated graphic
 would look like at time~$t$. Since the computation of the snapshot graphic is done
 by \TeX\ and since no  animation code is inserted into the output, this
 method works with arbitrary output formats, including~\textsc{pdf}.

% \begin{lstlisting}[style=example]
% \foreach \time in {1, 2, 3, 4} 
%   \tikz [make snapshot of=\time]
%     \graph [animated binary tree layout] { ... };
% \end{lstlisting}

% %\begin{center}
% %  \begin{lstlisting}[style=example]
% \noindent\hbox to\textwidth{\hss\foreach \time in {1, 2, 3, 4} {
%   \tikz [make snapshot of=\time, graph]
%     \graph [animated binary tree layout, nodes={fill=white}, fadein
%     time=200ms, fadeout time=200ms] {
%     { [when = 1] 10 -> { 5 -> { 2              , 7 -> 6  }, 15 -> 12   } },
%     { [when = 2] 10 -> { 5 -> { 2              , 7 -> 6  }, 15         } },
%     { [when = 3] 10 -> { 5 -> { 2              , 7       }, 15         } },
%     { [when = 4] 10 -> { 5 -> { 2 -> 3[second] , 7       }, 15         } }}; \qquad }\hss}
% %\end{lstlisting}
% %\end{center}

\section{Algorithmic Aspects of Drawing Dynamic Trees}
\label{section-algo}

Given a dynamic tree $T = (T_1,\dots,T_k)$ consisting of a sequence of
trees $T_i = (V_i, E_i, \phi_i)$, we saw in the introduction that
neither drawing each tree independently and then ``morphing'' the
subsequent drawings to create an animation nor laying out just the
supergraph  $\operatorname{super}(T) = (\bigcup_i V_i, \bigcup_i E_i)$
and then animating just the opacity of the nodes and edges will lead
to satisfactory drawings of dynamic trees. Our aim is to devise a new
algorithm that addresses the shortcomings of these approaches and that
meets a number of sensible \emph{aesthetic criteria} that we formulate
in Section~\ref{section-aesthetics}. The algorithm, presented in
Section~\ref{section-algorithm}, has
been implemented as a prototype~\cite{Skambath2016} and we have used
it to create the animations of dynamic trees in the present
paper. While the prototype implementation does not even run in linear
time (as would be possible by Theorem~\ref{theorem-acyclic}), it only
needs fractions of a second for the example graphs from this paper.

\subsection{Aesthetic Criteria for Drawing Dynamic Trees}
\label{section-aesthetics}

Already in 1979, Wetherell and Shannon~\cite{WetherellS1979}
explicitly defined aesthetic criteria for the layout of trees. Two
years later Reingold and Tilford~\cite{ReingoldT1981} 
refined these \emph{static criteria} towards more symmetric drawings
in which isomorphic subtrees must have the same layout. While the
criteria were originally formulated for binary trees only, one can
allow any number of children when there is an ordering on the
children of each node.

\begin{criterion*}[Ranking]\label{crit-ranking}
  The vertical position of a node equals its
  topological distance from the root. 
\end{criterion*}

\begin{criterion*}[Ordering]\label{crit-ordering}
  The horizontal positions of a node's children respect their
  topological order in the tree. 
\end{criterion*}

\begin{criterion*}[Centering]\label{crit-centering}
  Nodes are horizontally centered between their leftmost and rightmost
  child if there are at least two children.
\end{criterion*}

\begin{criterion*}[Symmetry]\label{crit-symmetry}
  All topologically order-isomorphic subtrees are drawn
  identically. Topologically mirrored subtrees are drawn horizontally 
  mirrored. 
\end{criterion*}

As numerous applications show, these rather sensible criteria lead to
aesthetically pleasing drawings of static trees. We extend these
criteria to the dynamic case. Ideally, we would like to keep all of the
above criteria, but will see in a moment that this is not always
possible.

\begin{wrapfigure}[6]{r}{0cm}
  \setbox\mybox=\hbox{
  \tikz {
    \scoped [graph, xshift=0cm]
       \graph [binary tree layout, significant sep=0pt] {
         a2/$n$["$T_i\colon$" left] -> { / -> {,/} , /$c$ }
       };
     \scoped [graph, xshift=2.5cm]
       \graph [binary tree layout, significant sep=0pt] {
         a3/$n$["$T_{i+1}\colon$" left] -> { / -> {,/} , /$c$ -> /}
       };
     }}
   \mydimen=\ht\mybox
   \advance\mydimen by-1.75em
   \ht\mybox=\mydimen
   \box\mybox
\end{wrapfigure}
Our first dynamic criterion forbids the unnecessary movement of
nodes in drawings like the one shown on the right, which shows the
same problem as the example in the introduction did: The horizontal
offset between $n$ and~$c$ changes from $T_i$ to $T_{i+1}$ even though
there is no structural change at~$n$. (Note that when a node disappears
in the step from $T_i$ to $T_{i+1}$ and then reappears in $T_{i+2}$,
the stability criterion does no require it to appear at the same
position as before.) 

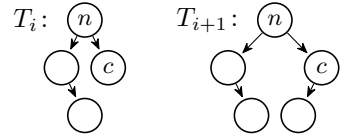
\begin{figure}[tb]
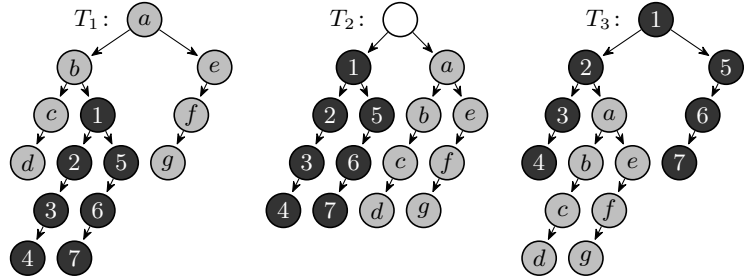

  \hfil
\tikz [graph, baseline] 
  \graph [binary tree layout, math nodes, significant sep=0pt] {  
    { [when = 1] a["$T_1\colon$" left,group 2]  -> { b[group 2] -> {
          { [nodes=group 2] c -> d , {[nodes=group 1] 1 -> { 2 ->
                3 -> 4 , 5->6 -> 7   } }}}, {[nodes = group 2]
            e -> f -> g  } }}};
\quad
\tikz [graph, baseline] 
  \graph [binary tree layout, math nodes, significant sep=0pt] {
    { [when = 2] /["$T_2\colon$" left] -> { {[nodes=group 1] 1 -> { 2
            -> 3 -> 4,  5 -> 6 -> 7  }}, { [nodes=group 2] a
          -> { b -> c -> d , e -> f -> g  }}}}};
\quad
\tikz [graph, baseline] 
  \graph [binary tree layout, math nodes, significant sep=0pt] {  
    { [when = 3] 1["$T_3\colon$" left,group 1]  -> { 2[group 1] -> {
          { [nodes=group 1] 3 -> 4 , {[nodes=group 2] a -> { b ->
                c -> d , e->f -> g   } }}}, {[nodes = group 1]
            5 -> 6 -> 7  } }}};
  \caption{A ``problematic'' dynamic tree. Already the dynamic tree $T
    = (T_1,T_2)$ cannot be drawn while meeting all of the criteria
    Ranking, Ordering, Centering, Symmetry, and Stability, as shown in 
    Lemma~\ref{lemma-impossible}. The whole dynamic tree $T =
    (T_1,T_2,T_3)$ cannot even be drawn when the Symmetry Criterion is
    replaced by the Weak Symmetry Criterion, see
    Lemma~\ref{lemma-still-impossible}.} 
  \label{figure-problem}  
\end{figure}

\begin{criterion*}[Stability]\label{crit-stability}
  The horizontal offset between a node~$n$ and a child~$c$ may 
  not change between the layouts of trees $T_i$ and $T_{i+1}$ if $c$
  does not change its position in the ordering of the children of~$n$.
\end{criterion*}

While the stability criterion forbids relative movements of
connected nodes, it allows whole subtrees to move without changing their inner
layout. This emphasizes the important parts of changes since multiple
objects moving with the same speed are percieved as one
connected group \cite{BartramW2002,WareB2004}.
The criterion reduces movements and draws common structures
identically, thereby reducing errors in understanding
\cite{ArchambaultPP2011} and making it easier for viewers to correctly
recognize the changes in the tree sequence~\cite{ArchambaultP2012}. % -- even for small multiples instead of 
%animations.% recognising changes might get easier when common structures
%are drawn the 
%same.  

While all of the above criteria are reasonable, unfortunately, there
is no way of meeting all of them simultaneously, see the appendix for
the proof:

\begin{lemma}\label{lemma-impossible}
  No drawing of the dynamic tree $T = (T_1,T_2)$ from
  Figure~\ref{figure-problem} meets all of the
  criteria Ranking, Ordering, Centering, Symmetry, and Stability. 
\end{lemma}

% \begin{proof}[Sketch of Proof]
%   Use $T_1$ and $T_2$ from Figure~\ref{figure-problem}. The two
%   order-isomorphic subtrees in $T_2$ (the nodes with numbers and the
%   nodes with letters) must be drawn identically by Symmetry. By
%   Stability, the distances between $2$ and $5$ in $T_1$ and $T_2$ must
%   be identical, just like the distance between $b$ and $e$ in both
%   trees. By Ranking and Ordering, the nodes $d,2,5,g$ must come in
%   this order on a line in $T_1$. However, now, the distance between
%   $d$ and $g$ must be greater than the distance between $2$ and~$5$, a
%   contradiction. \qed
% \end{proof}

In view of the lemma, we will need to weaken one or more of our
criteria, while still trying to meet them at least in ``less
problematic'' cases than the dynamic tree from
Figure~\ref{figure-problem}. Furthermore, even when the criteria
\emph{can} be met, this may not always be desirable.

\begin{wrapfigure}[8]{r}{0cm}
  \setbox\mybox=\hbox{%
    \tikz [graph]{
      % \useasboundingbox [] (-1,-0.3) rectangle (4,-2.25);
      \graph [binary tree layout, math nodes, significant sep=0pt] {  
        { [when = 1] /["$T_1\colon$" left]  -> {
            { [nodes=group 1]1 -> { 2, 3 }},
            / -> { [nodes=group 2] a[second] -> { b ->[white] /[white,second], c
                ->[white] /[white] }
            }
          }
        }};
      \scoped [xshift=2.5cm]
      \graph [binary tree layout, math nodes, significant sep=0pt] {  
        { [when = 2] /["$T_2\colon$" left]  -> {
            { [nodes=group 1]1 -> { 2, 3 }},
            / -> { a[second, group 2] -> { b[group 2] ->[] /[second],
                c[group 2]
                ->[] /[] }
            }
          }
        }};
    }
  }
  \mydimen=\ht\mybox
  \advance\mydimen by-2.5em
  \ht\mybox=\mydimen
  \box\mybox
\end{wrapfigure}
Consider the right example, which seems like a ``reasonable'' drawing of a
dynamic tree. The Stability Criterion enforces the large distance 
between $b$ and $c$ already in $T_1$, but the Symmetry Criterion would
now actually enforce the same distance between $2$ and~$3$, which
seems undesirable here. As a replacement of the Symmetry
Criterion we propose a \emph{Weak Symmetry
  Criterion} that our algorithm will be able to meet in many important
cases, including the troublesome example from 
Lemma~\ref{lemma-impossible}. Nevertheless, there are \emph{still}
dynamic trees that cannot be drawn in this way, see
Lemma~\ref{lemma-still-impossible}, which also turn out to be the
algorithmically difficult cases.

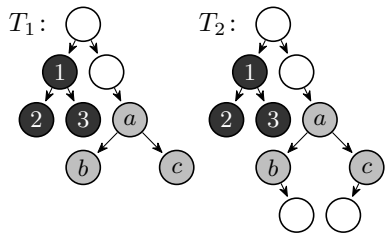
\begin{criterion*}[Weak Symmetry]
  Let $n$ and $n'$ be nodes such that for all $i\in\{1,\dots,n\}$ the
  subtrees rooted at $n$ and at $n'$ in $T_i$ are order-isomorphic
  (or all mirrored). Then in all drawings of the~$T_i$ the subtrees
  rooted at $n$ and $n'$ must all be drawn identically (or all
  mirrored).  
\end{criterion*}

\subsection{An Algorithm for Drawing Arbitrary Dynamic Trees}
\label{section-algorithm}

Our starting point for an algorithm that meets the aesthetic criteria
just formulated is the classical Reingold--Tilford
algorithm~\cite{ReingoldT1981}. It will be useful to review this
algorithm briefly, formulated in a ``bottom-up'' fashion: While there
is a node that has not yet been processed, pick a node $n$ whose
children $c^1,\dots,c^m$ have all already been processed (this is
immediately the case for all leafs, where $m=0$). For each child $c^r$
a layout $L(c^r)$ will have been computed for the subtree $T(c^r)$
of~$T$ rooted at~$c^r$. The algorithm now shifts the $L(c^r)$
vertically so that all $c^r$ lie 
on the same horizontal line (Ranking Criterion),
then shifts them horizontally so that the $c^1$ comes first, followed
by $c^2$, and so on (Ordering Criterion), such that
no overlap of the $L(c^r)$ occurs. Finally, $n$ is centered above its
children (Centering Criterion). The Symmetry Criterion is satisfied 
automatically by this algorithm since the same shifts occur for
symmetric subtrees. Using appropriate data structures, the algorithm
can be implemented in linear time.

Our Algorithm~\ref{algo-acyclic}, see the appendix for pseudo-code, uses
the same basic idea as the Reingold--Tilford algorithm, but introduces
two new ideas. 

\begin{wrapfigure}[8]{r}{0cm}
  \tikz[yscale=.6,xscale=.9, group 1/.append style={draw=white}]{
    \begin{scope}[graph,nodes=node]
%    \useasboundingbox (0,0) rectangle (3,2) [draw];
  \begin{scope}[name=ba, base=80, fill=white]
    \filldraw[group 1=80] (-1,-1) --  +(-.5,-2) --+(.5,-2) -- cycle;
    \filldraw[group 2=80] (+1,-1) -- + (-.25,-1) -- +(-.125,-1.5)
-- +(-.25,-2) -- +(.5,-2) -- cycle;
    \node[fill,] (a1) at (0,.3) {};
    \node[fill, group 1=80] (c) at (-1,-1) {};
    \node[fill, group 2=80] (d) at (1,-1) {};
    \draw[-,] (a1) -- (c);
    \draw[-,] (a1) -- (d);
  \end{scope}
  \begin{scope}[name=bb,xshift=-.2cm,yshift=-.15cm,base=90,fill=white]
    \filldraw[group 1=90] (-1,-1) --  +(-.5,-2) -- +(.25,-2) --+(.4,-1.5) -- cycle;
    \filldraw[group 2=90] (+1,-1) -- + (-.25,-1) -- +(-.125,-1.5)
    -- +(-.25,-2) -- +(.5,-2) -- cycle;
    \node[fill,] (a) at (0,.3) {};
    \node[fill, group 1=90] (c) at (-1,-1) {};
    \node[fill, group 2=90] (d) at (1,-1) {};
    \draw[-,] (a) -- (c);
    \draw[-,] (a) -- (d);
  \end{scope}
  \begin{scope}[name=bc,xshift=-.4cm,yshift=-.3cm,fill=white,base=100]
    \filldraw[group 1] (-1,-1) --  +(-.5,-2) -- +(.125,-2) -- +(0,-1.5) --+(.4,-1.5) -- cycle;
    \filldraw[group 2] (+1,-1) -- + (-.25,-1) -- +(-.125,-1.5)
-- +(-.25,-2) -- +(.5,-2) -- cycle;
    \node[fill,] (a) at (0,.3) {};
    \node[fill,group 1] (c) at (-1,-1) {};
    \node[fill,group 2] (d) at (1,-1) {};
    \draw[-,] (a) -- (c);
    \draw[-,] (a) -- (d);
  \end{scope}
  \begin{scope}[xshift=3cm,name=bb]
      \begin{scope}[name=ba, base=80]
    \filldraw[group 1=80] (-.3,-1) --  +(-.5,-2) --+(.5,-2) -- cycle;
    \filldraw[group 2=80] (+.3,-1) -- + (-.25,-1) -- +(-.125,-1.5)
-- +(-.25,-2) -- +(.5,-2) -- cycle;
    \node[fill,] (a2) at (0,.3) {};
    \node[fill, group 1=80] (c) at (-.3,-1) {};
    \node[fill, group 2=80] (d) at (.3,-1) {};
    \draw[-,] (a2) -- (c);
    \draw[-,] (a2) -- (d);
  \end{scope}
  \begin{scope}[name=bb,xshift=-.2cm,yshift=-.15cm,base=90]
    \filldraw[group 1=90] (-.3,-1) --  +(-.5,-2) -- +(.25,-2) --+(.4,-1.5) -- cycle;
    \filldraw[group 2=90] (+.3,-1) -- + (-.25,-1) -- +(-.125,-1.5)
    -- +(-.25,-2) -- +(.5,-2) -- cycle;
    \node[fill,] (a) at (0,.3) {};
    \node[fill,group 1=90] (c) at (-.3,-1) {};
    \node[fill,group 2=90] (d) at (.3,-1) {};
    \draw[-,] (a) -- (c);
    \draw[-,] (a) -- (d);
  \end{scope}
  \begin{scope}[name=bc,xshift=-.4cm,yshift=-.3cm,base=100]
    \filldraw[group 1] (-.3,-1) --  +(-.5,-2) -- +(.125,-2) -- +(0,-1.5) --+(.4,-1.5) -- cycle;
    \filldraw[group 2] (+.3,-1) -- + (-.25,-1) -- +(-.125,-1.5)
-- +(-.25,-2) -- +(.5,-2) -- cycle;
    \node[fill,] (a) at (0,.3) {};
    \node[fill, group 1] (c) at (-.3,-1) {};
    \node[fill, group 2] (d) at (.3,-1) {};
    \draw[-,] (a) -- (c);
    \draw[-,] (a) -- (d);
  \end{scope}
\end{scope}
\end{scope}

\draw [, >={Stealth[round]}, semithick, bend right=10, dashed]
    (-.3,-3.5) edge [->,"tighten"'] (2.7,-3.5);
}
\end{wrapfigure}
\emph{Idea 1: Treat Nodes as Three-Dimen\-sio\-nal Objects.} In our
algorithm, we treat nodes and subtrees as ``three dimensional''
objects with time as the third dimension. Given a dynamic tree $T =
(T_1,\dots,T_k)$, the algorithm does not process the $T_i$ one at a
time (as online algorithms have to do), but instead considers
for each node~$n$ of the supergraph $\operatorname{super}(T)$ the
sequence $(T_1(n),\dots,T_k(n))$ of trees rooted at~$n$ in the
different~$T_i$ and computes a whole sequence of layouts
$(L_1(n),\dots,L_k(n))$ for these trees: 
The core operation of the Reingold--Tilford algorithm, the shifting of a
layout $L(c^r)$ until it almost touches the previous layout $L(c^{r-1})$, is
replaced by a shifting of the whole sequence $(L_1(c^r_1),\dots,
L_k(c^r_k))$, where $\smash{c_j^{i}}$ denotes the $i$th child of~$n$ in
$T_j$, until at least one layout $L_j(\smash{c^r_j})$ (one of the 
gray layouts in the example)  almost touches its sibling's layout
$L_j(c^{r-1}_j)$ (one of the dark layouts).   

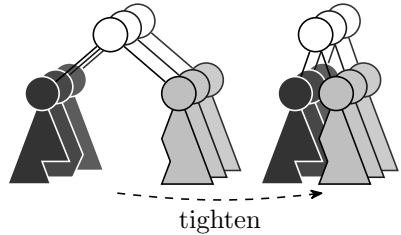
\emph{Idea 2: Processing the Supergraph Using a Topological Ordering.}
For static trees, there is a clear order in which the nodes should be
processed by the Reingold--Tilford algorithm: from the leafs
upwards. For a dynamic tree, this order is no longer clear -- just
consider the example from Figure~\ref{figure-problem}: Should we first
process node~$1$ or node~$a$?
Our algorithm address this ordering problem as follows: We compute the
supergraph $\operatorname{super}(T)$ and then check whether it is
acyclic. If so, it computes a topological ordering of
$\operatorname{super}(T)$ and then processes the nodes in this
order. Observe that this guarantees that whenever a node is processed,
complete layouts for its children will already have been
computed. 

\begin{theorem}\label{theorem-acyclic}
  Let $T$ be a dynamic tree whose supergraph is acyclic. Then
  Algorithm~\ref{algo-acyclic} draws $T$ in linear time such that all of the criteria
  Ranking, Ordering, Centering, Weak Symmetry, and Stability are met.  
\end{theorem}

%\subsection{Drawing Dynamic Trees with Cyclic Supergraphs\\ Optimally
%  is NP-Complete} 
%\label{section-cycleremoval}

Theorem~\ref{theorem-acyclic} settles the problem of drawing dynamic
trees with acyclic supergraphs nicely. In contrast, for a cyclic
supergraph, things get \emph{much} harder:

\begin{lemma}\label{lemma-still-impossible}
  No drawing of $T = (T_1,T_2,T_3)$ from
  Figure~\ref{figure-problem} meets all of the
  criteria Ranking, Ordering, Centering, Weak Symmetry, and Stability. 
%
%  The dynamic tree from Figure~\ref{figure-problem} cannot be drawn
%  while meeting all of the criteria Ranking, Ordering, Centering, Weak
%  Symmetry, and Stability.  
\end{lemma}

The lemma tempts us to just ``give up'' on cyclic
supergraphs. However, these arise naturally in prune-and-regraft
operations and from rotations in search 
trees -- which are operations that we would like to visualize. We
could also just completely ignore the temporal criteria and return to
drawing each tree individually in such cases -- but we might be able
to draw everything nicely except for a single ``small'' cycle 
``somewhere'' in the supergraph. 

We propose to deal with the cycle problem by ``cutting'' the cycles
with as few ``temporal cuts'' as possible. These are defined as
follows: Let $G = (G_1,\dots,G_k)$ be a dynamic graph and let $n$ be a
node of the supergraph $\operatorname{super}(G)$ and let $i \in
\{1,\dots,k-1\}$. The \emph{temporal cut} of $G$ at $n$ and $i$ is a new
dynamic graph $G'$ that is identical to $G$, except that for all $j
\in \{i+1,\dots, k\}$ in which $G_j$ contains the node $n$, this node
is replaced by the same new node~$n'$ (and all edges to or from $n$
are replaced by edges to or from $n'$).

Temporal cuts can be used to remove cycles from the supergraph of a
dynamic graph, which allows us to then run our
Algorithm~\ref{algo-acyclic} on the resulting graph; indeed, simply
``cutting everything  at all times'' turns every supergraph into a
(clearly acyclic) collection of non-adjacent edges and isolated
nodes. However, we wish to minimize the number of temporal cuts since,
when we visualize $G'$ using an animation, the different locations
that may be assigned to $n$ and $n'$ will result in a \emph{movement}
of the node $n$ to the new position of~$n'$.

By the above discussion, we would like to find an algorithm that
solves the following problem $\Lang{temporal-cut-minimization}$: Given
a dynamic tree $T$, find a minimal number of temporal cuts, such that
the resulting dynamic tree $T'$ has an acyclic
supergraph. Unfortunately, this problem turns out to be difficult:

\begin{theorem}\label{theorem-np}
  The decision version of $\Lang{temporal-cut-minimization}$ is
  $\Class{NP}$-complete. 
\end{theorem}

In light of the above theorem, we have developed and implemented a
simple greedy heuristic, Algorithm~\ref{algo-greedy}, for finding 
temporal cuts that make the supergraph acyclic, which our prototype
runs prior to invoking Algorithm~\ref{algo-acyclic}: Given a dynamic
tree, the heuristic simply adds the trees~$T_i$ and their edges
incrementally to the supergraph. However, whenever adding an edge
$e=(v,w)$ of~$T_i$ to the supergraph creates a cycle, we cut $w$
at $i-1$.  

\section{Conclusion and Outlook}

\begin{wrapfigure}[7]{r}{0cm}
  \linktosvg{
  \setbox\mybox=\hbox{%
  \tikz [small graph, bg=10] 
  \graph [animated binary tree layout, empty nodes] {  
    { [when = 0s] a[group 2]  -> { b[group 2] -> {
          { [nodes=group 2] c -> d , {[nodes=group 1] 1 -> { 2 ->
                3 -> 4 , 5->6 -> 7   } }}}, {[nodes = group 2]
            e -> f -> g  } }},
    { [when = 3s] /[] -> { {[nodes=group 1] 1 -> { 2
            -> 3 -> 4,  5 -> 6 -> 7  }}, { [nodes=group 2] a
          -> { b -> c -> d , e -> f -> g  }}}},
    { [when = 6s] 1[group 1]  -> { 2[group 1] -> {
          { [nodes=group 1] 3 -> 4 , {[nodes=group 2] a -> { b ->
                c -> d , e->f -> g   } }}}, {[nodes = group 1]
          5 -> 6 -> 7  } }}};}
   \mydimen=\ht\mybox
   \advance\mydimen by-2.5em
   \ht\mybox=\mydimen
   \box\mybox}
\end{wrapfigure}\looseness=-1
We have presented a system for offline drawings of dynamic trees using
animations that are embedded in (text) documents. The system has been
implemented~\cite{Skambath2016} as an extension of the popular \TeX\
system and will become part of future version of
\TikZ.\footnote{Currently available in the development version at
  \url{http://pgf.cvs.sourceforge.net}.} The generated 
animation are light-weight both in terms of file size and generation
time, but require that the documents (or, at least, the graphic files)
are stored in the \textsc{svg} format. Our new algorithm is a natural
extension of the Reingold--Tilford algorithm to the dynamic case, but
while the original algorithm runs in linear 
time on all trees, we showed that the dynamic case leads to
$\Class{NP}$-complete problems. Fortunately, in practice, the hard
subproblems can be solved satisfactorily using a greedy strategy -- at
least, that has been our finding for a limited number of examples such
as the above animation; a perceptual study of animated drawings of
dynamic graphs has not (yet) been conducted. 

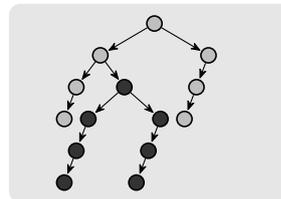
\looseness=-1
We see our algorithm as a first step towards a general set of
algorithms for drawing dynamic graphs using animations, which we
believe to have a great (and not yet fully realized) potential as
parts of text documents. A next logical step would be a transferal of
the Sugiyama method~\cite{EadesS1990,SugiyamaTT1979} to the dynamic
offline case.

\bibliographystyle{plain}
\bibliography{main}

\clearpage
\appendix
\section{Technical Appendix}

In this appendix, Section~\ref{section-gallery} presents some example
animations for dynamic trees that were generated using our prototype
implementation. In Section~\ref{section-algos} we present pseudo-code
for our two algorithms: The main Algorithm~\ref{algo-acyclic} for
drawing dynamic graphs with an acyclic supergraph, and the greedy
heuristic  Algorithm~\ref{algo-greedy} for finding temporal cuts
quickly. Finally, Section~\ref{section-proofs} contains the proofs for
the lemmas and theorems from the main text.

\subsection{A Gallery of Animations of Dynamic Trees}
\label{section-gallery}

In the following, we present a number of animations that show how
different dynamic graphs evolve over time. These animations will play
only in the \textsc{svg} version of this
paper~\cite{SkambathT2016-svg}, the \textsc{pdf} version displays only
the (rather boring) initial tree $T_1$ of the dynamic tree. As
explained in Section~\ref{section-snapshots}, it would have been
possible to generate snapshots of the animations, but since the whole
point of this paper is to show animations embedded in documents, we
have only include the animations.

\begin{wrapfigure}[11]{r}{0cm}
  \def\speed{2s}
%\linktosvg{
  \setbox\mybox=\hbox{\linktosvg{%
  \tikz[graph,bg=10]{
        \path (-3.5,-3.5) (3.5,0.25);
        \scoped[overlay]
  \graph [animated binary tree layout, math nodes] {  
    { [when = 0s] a[group 2]  -> { b[group 2] -> {
          { [nodes=group 2] c -> d , {[nodes=group 1] 1 -> { 2 ->
                3 -> 4 , 5->6 -> 7   } }}}, {[nodes = group 2]
            e -> f -> g  } }},
    { [when = 3s] /[] -> { {[nodes=group 1] 1 -> { 2
            -> 3 -> 4,  5 -> 6 -> 7  }}, { [nodes=group 2] a
          -> { b -> c -> d , e -> f -> g  }}}},
    { [when = 6s] 1[group 1]  -> { 2[group 1] -> {
          { [nodes=group 1] 3 -> 4 , {[nodes=group 2] a -> { b ->
                c -> d , e->f -> g   } }}}, {[nodes = group 1]
          5 -> 6 -> 7  } }}};}}}
   \mydimen=\ht\mybox
   \advance\mydimen by-1.5em
   \ht\mybox=\mydimen
   \box\mybox
%}
\end{wrapfigure}
Our first example is the ``troublesome'' dynamic tree from
Figure~\ref{figure-problem} as rendered by our algorithm. As we argued
in detail in Lemma~\ref{lemma-still-impossible}, it is not possible to
draw this dynamic tree without violating the Stability Criterion at
least once. In the drawing in Figure~\ref{figure-problem}, this
criterion is actually violated twice, namely for
the node~$a$ between $T_1$ and $T_2$ and again for the node~$1$
between $T_2$ and $T_3$. In contrast, our algorithm, which tries to
minimize these violations, succeeds in only having to ``cut once''
and, hence, renders the dynamic tree with a single ``movement''
resulting from this cut.

\begin{wrapfigure}[11]{r}{0cm}
\def\speed{2s}
  \setbox\mybox=\hbox{\linktosvg{%
      \tikz[large graph,bg=10]{
        \path (-3.5,-3.5) (3.5,0.3);
        \scoped[overlay]
  \graph[animated binary tree layout,
         split critical arc head=false,
         split critical arc tail=true,
   ] {
   {[when=\speed*0] 45 },
   {[when=\speed*1] 45->{36,} },
   {[when=\speed*2] 36->{12,45} },
   {[when=\speed*3] 36->{12->{,28},45} },
   {[when=\speed*4] 36->{12->{0,28},45} },
   {[when=\speed*5] 28->{12->{0,13},36->{,45}} },
   {[when=\speed*6] 28->{12->{0,13},45->{36,49}} },
   {[when=\speed*7] 28->{12->{0,13->{,21}},45->{36,49}} },
   {[when=\speed*8] 28->{12->{0,21->{13,23}},45->{36,49}} },
   {[when=\speed*9] 28->{13->{12->{0,},21->{20,23}},45->{36,49}} },
   {[when=\speed*10] 28->{13->{12->{0,},21->{20,23}},45->{,49}} },
   {[when=\speed*11] 28->{13->{12->{0,},21->{20,23}},45->{30,49}} },
   {[when=\speed*12] 28->{13->{12->{0,},21->{,23}},45->{30,49}} },
   {[when=\speed*13] 28->{13->{7->{0,12},21->{,23}},45->{30,49}} },
   {[when=\speed*14] 13->{7->{0,12->{11,}},28->{21->{,23},45->{30,49}}} },
   {[when=\speed*15] 13->{7->{0,12->{11,}},28->{21->{14,23},45->{30,49}}} },
   {[when=\speed*16] 13->{7->{0,12->{11,}},28->{21->{14,23->{22,}},45->{30,49}}} },
   {[when=\speed*17] 21->{13->{11->{7,12},14},28->{23->{22,},45->{30,49}}} },
   {[when=\speed*18] 21->{11->{7->{3,},13->{12,14}},28->{23->{22,},45->{30,49}}} },
   {[when=\speed*19] 21->{11->{7->{3,},13->{12,}},28->{23->{22,},45->{30,49}}} },
   {[when=\speed*20] 21->{11->{7->{3,},13->{12,}},28->{23,45->{30,49}}} },
   {[when=\speed*21] 21->{11->{7->{3,},13->{12,}},28->{23->{22,},45->{30,49}}} },
   {[when=\speed*22] 21->{12->{7->{3,},13},28->{23->{22,},45->{30,49}}} },
   {[when=\speed*23] 21->{12->{7->{3,},13},28->{23->{22,},45->{,49}}} },
   };
 }}}
   \mydimen=\ht\mybox
   \advance\mydimen by-1.5em
   \ht\mybox=\mydimen
   \box\mybox
\end{wrapfigure}
Our second example is a rendering of an \textsc{avl} tree that gets
updated repeatedly, namely by 24 update operations, consisting both of
insertions and deletions. The \textsc{avl} tree is balanced, meaning that
rotations are used to keep the height of the tree logarithmically
bounded. Each rotation introduces a cyclic dependency in the
supergraph, all of which have to be removed by
Algorithm~\ref{algo-greedy} and all of which result in movements in
the animation. While the Stability Criterion dictates that such
movements should not happen, the animation shows that -- besides being
unavoidable -- they are actually helpful in the example since they
highlight the places where rotations occur.

\begin{wrapfigure}[10]{r}{0cm} % Anzahl Zeilen, kurz sein sollen, in
                           % eckigen Klammern so wie in den oberen Beispielen
\def\speed{2s}
\linktosvg{
  \setbox\mybox=\hbox{%
\tikz[large graph,bg=10]{
  \graph[animated binary tree layout, math nodes,
         split critical arc head=false,
         split critical arc tail=true,
         ] {  
           {[when=\speed*0] 20->{8->{,15},24->{,30}} },
           {[when=\speed*1] 20->{8->{,15->{,18}},24->{,30}} },
           {[when=\speed*2] 20->{8->{,15->{,18}},24->{,30}} },
           {[when=\speed*3] 20->{8->{4,15->{,18}},24->{,30}} },
           {[when=\speed*4] 20->{8->{4,15->{,18}},24->{23,30}} },
           {[when=\speed*5] 20->{8->{4,15->{,18}},24->{23,30->{26}}} },
           {[when=\speed*6] 20->{15->{4,18},24->{23,30->{26}}} },
           {[when=\speed*7] 20->{15->{4,18},24->{23->{21},30->{26}}} },
           {[when=\speed*8] 20->{15->{4,18},23->{21,30->{26}}} },
           {[when=\speed*9] 20->{15->{4,18},23->{21,30->{26,34}}} },
           {[when=\speed*10] 20->{15->{4,18},21->{,30->{26,34}}} },
           {[when=\speed*11] 20->{15->{4,18},21->{,30->{,34}}} },
           {[when=\speed*12] 20->{15->{4,18},30->{,34}} },
           {[when=\speed*13] 20->{15->{4,18},30->{27,34}} },
           {[when=\speed*14] 20->{15->{4,18},30->{27,34}} },
           {[when=\speed*15] 20->{15->{4,18->{,19}},30->{27,34}} },
           {[when=\speed*16] 20->{15->{4,18->{,19}},30->{27->{,28},34}} },
           {[when=\speed*17] 20->{15->{4,18->{,19}},30->{27->{,28},34->{32}}} },
           {[when=\speed*18] 20->{15->{4->{,10},18->{,19}},30->{27->{,28},34->{32}}} },
         };}}
   \mydimen=\ht\mybox
   \advance\mydimen by-1.5em
   \ht\mybox=\mydimen
   \box\mybox
}
\end{wrapfigure}
As a final example we have a rendering of a simple non-balanced binary
search tree (20 steps). In this search tree new nodes are inserted as
new leafs of the tree and in this example each node that is deleted
has only one single child, which takes its parent's position. In
contrast to the previous example of an \textsc{avl} tree, neither a
rotation occurs nor does the ascendant-descendant relation change for any
pair of nodes from one tree~$T_i$ to the next. Hence, the supergraph
is acyclic and the produced animation meets all of the criteria Ranking,
Ordering, Centering, Stability, and Weak Symmetry.

\subsection{Algorithms in Pseudo-Code}
\label{section-algos}

\begin{algorithm}[Drawing Dynamic Trees with Acyclic Supergraphs]
    \label{algo-acyclic}
    \hfil
\begin{lstlisting}[style=pseudocode]
input a dynamic tree $T = (T_1,\dots,T_k)$

compute $\operatorname{super}(T)$

// Make the supergraph acyclic
if $\operatorname{super}(T)$ is not acyclic then
    call Algorithm$~\ref{algo-greedy}$
    update $T$ and recompute $\operatorname{super}(T)$, which is now acyclic

sort the vertices of $\operatorname{super}(T)$ topologically so that $\{v_1,\dots,v_n\}$ is the vertex set of $\operatorname{super}(T)$ and $\text{for}$ all edges $(v_i,v_j)$ we have $j<i$

// Iterate over all nodes
for $i \gets 1$ to $n$ do
    $m \gets$ the maximum number of children $v_i$ has over time

    // By the sorting, for every child $c_j^r$ of $v_i$, the layout $L_r(c_j^r)$ will already have been computed
    for $r \gets 2$ to $m$ do 
        foreach $T_j$ that contains $v_i$ do
            // Use the Reingold-Tilford data structure to compute in linear time:
            $\operatorname{dist}^r_j(v_i) \gets$ minimum horizontal distance from $c^{r-1}_j$ $\text{to}$ $c^r_j$ so that all of $L_j(c^{1}_j),\ldots,L_j(c^{r-1}_j)$ is $\text{to}$ the left of $L_j(c^{r}_j)$ with a fixed minimal padding
    
        // Synchronize the distance between neigboring subtrees
        $\operatorname{dist}^{r}(v_i) \gets \max_j \{\operatorname{dist}^r_j(v_i)\}$ 

        // Shift the subtree
        foreach $T_j$ that contains $v_i$ do
            $x_j(c^r_j)\gets x_j(c^{r-1}_j)+\operatorname{dist}^{r}(v_i)$ 
            update the data structure of Reingold and Tilford $\text{for}$ the used shift
        
    // Update the relative shift such that $L(v_i)$ is centered:
    $\operatorname{width}(v_i)\gets \sum_{r=2}^m \operatorname{dist}^r(v_i)$
    foreach $T_j$ that contains $v_i$ do
        $x_j(v_i)\gets 0$ // the initial horizontal shift of $v_i \text{ in }L_j(v_i)$
        for $r\gets 1$ to $m$ do
            $x_j(c^r_j)\gets x_j(c^r_j)-\frac{1}{2}\operatorname{width}(v_i)$

// Compute absolute coordinates
foreach snapshot $T_i$ do
    compute the depth of each node $v\text{ in }T_i$ as the vertical coordinate of $L_i(v)$ 
    compute the horizontal position of all nodes by accumulation of all shift values on the path from the root $\text{to}$ the node $\text{in}$ a tree traversal.

return $(L_1,\dots,L_k)$
\end{lstlisting}
\end{algorithm}

\begin{algorithm}[Greedy Heuristic for Making Supergraphs Acyclic]
  \label{algo-greedy}
  \hfil
\begin{lstlisting}[style=pseudocode]
input a dynamic tree $T = (T_1,\dots,T_k)$

$V' \gets \emptyset$
$E' \gets \emptyset$

for $i \gets 1$ to $k$ do
    let $T'_i=(V'_i,E'_i)$ be a new tree with $V'_i=E'_i=\emptyset$

    foreach edge $(v,w)\in E_i$ do
        // Check if the edge creates a cycle in the supergraph
        if $v$ is reachable from $w \text{ in } (V'\cup V'_i,E'\cup E'_i)$ then
            replace $w$ by $w'\text{ in}$ all trees $T_j$ with $j\geq i$ and $\text{in }T'_i$
            add the edge $(v,w')\text{ to }T'_i$ 
        else
            add the edge $(v,w)\text{ to }T'_i$

    // Add possibly renamed nodes and edges to the still-acyclic supergraph
    $V' \gets V' \cup V'_i$
    $E' \gets E' \cup E'_i$

return $T'=(T'_1,\dots,T'_k)$ 
\end{lstlisting}
\end{algorithm}

\subsection{Proofs Omitted from the Main Text}
\label{section-proofs}

\begin{proof}[of Lemma~\ref{lemma-impossible}]
  We use the first two trees $T_1$ and $T_2$ from
  Figure~\ref{figure-problem}. For a node~$x$ of a tree $T_i$, let us write
  $h^x_i$ for the horizontal distance from $x$ to the next node to the
  right of it on its height (for instance, $h^c_2$ is the distance from
  $c$ to $f$ in~$T_2$). In $T_2$, the subtrees rooted at~$1$ and
  at~$a$ are clearly order-isomorphic and, hence, have to be drawn identically by
  the Symmetry Criterion. The same is true for the subtrees rooted at
  $2$, $5$, $b$, and $e$. Hence, $h^2_2 = h^3_2 = h^4_2 = h^b_2 =
  h^c_2 = h^d_2$. 

  Consider $T_1$ and observe that \emph{as in $T_2$} the nodes
  $2$ and~$5$ are the first and second child of the node~$1$ and the
  nodes $b$ and~$e$ are the first and second of~$a$. By the Stability
  Criterion, we get $h^2_1 = h^2_2$ and $h^b_1 = h^b_2$. Now, in the
  tree $T_1$, by the Ranking and Ordering Criteria, the vertices
  $d,2,5,g$ must come in this order as shown. However, the distance
  between $d$ and $g$ in $T_1$, which must be equal to $h^b_1 =
  h^2_2$, is clearly greater than $h^2_1 = h^2_2$; and $h^b_2 > h^2_2$
  is a contradiction to $h^b_2 = h^2_2$.
  \qed 
\end{proof}

\begin{proof}[of Theorem~\ref{theorem-acyclic}]
  Algorithm~\ref{algo-acyclic} automatically meets the criteria 
  Ranking, Ordering, and Centering: As in independent runs of the
  Reingold--Tilford algorithm for each~$T_i$,  only the horizontal
  distance between neighboring children of the same node differs, which
  does not influence the ordering of children. Furthermore, as the
  algorithm processes all nodes in a topological order, the layout of
  a node~$n$ depends only on the previously computed subtree layouts
  $L_i(c^j_i)$ of $n$'s children and thus the Weak symmetry criterion
  holds automatically, too. 

%  Since the use of the data structure from the Reingold--Tilford algorithm just
%  improves the runtime, the computed layouts would be the same as if the 
%  algorithm computes and performs the required shift of neighbored
%  subtree layouts directly. Thereby, 
 
  Since processing a node~$n$ just shifts the children (with their
  subtrees) of a node~$n$ relative to~$n$ and with the same horizontal
  distances $\operatorname{dist}^r(n)$ between every $(r-1)$-th and
  $r$-th child in each~$T_i$, the produced layout meets the Stability
  criterion: A node only ``moves'' in two cases. First, its parent or
  its position relative to its siblings can change; but then the
  Stability Criterion makes no requirements. Second, there may be a
  change at some some ancestor further up; but then there is no
  change of the offset between $n$ and its parent node since the inner
  layout of the related subtree is already fixed.
    
  Concerning the claimed linear runtime, the only difficult part is to
  see how the computation of the necessary shifts can be done in
  linear time: A na\"\i ve implementation would remember and then
  traverse the ``left'' and ``right'' sides of the different layouts
  repeatedly to 
  compute the point of ``least distance'' between adjacent
  layouts. Reingold and Tilford had a clever idea of introducing a
  skipping data structure that removes this requirement: One can
  compute the necessary shift distance between two given layouts of
  subtrees in constant time using this data structure. This yields a
  linear runtime for the classical Reingold--Tilford algorithm and
  also a linear runtime for our algorithm since we can use the same
  data structure and need to compute the same shift distances.
  \qed
\end{proof}

\begin{proof}[of Lemma~\ref{lemma-still-impossible}]
  The tree $T = (T_1,T_2,T_3)$ from Figure~\ref{figure-problem} cannot
  be drawn while meeting the criteria. The argument is essentially the
  same as in Lemma~\ref{lemma-impossible}, only we can now no longer
  argue that in $T_2$ we must have $h^2_2 = h^2_b$ since the trees
  rooted at $a$ and $1$ no longer have the same overall ``history''
  and the Weak Symmetry Criterion does not apply to them. However, it
  \emph{does} apply to the trees rooted at $2$, $5$, $b$, and $e$ and,
  hence, we have $h^2_2 = h^3_2 = h^4_2$ and $h^b_2 = h^c_2 = h^d_2$.
  Furthermore, as in Lemma~\ref{lemma-impossible}, the Stability
  Criterion still tells us $h^2_1 = h^2_2 = h^2_3$ and $h^b_1 = h^b_2
  = h^b_3$. Finally, as in the Lemma~\ref{lemma-impossible}, the
  Ranking and Ordering Criteria still yield that in~$T_1$ the nodes
  $d,2,5,g$ must be ordered as shown in Figure~\ref{figure-problem}
  and in $T_3$ the nodes $4,b,e,t$ must be ordered as shown. From
  these orderings we can conclude $h^b_1 > h^2_1$ and $h^2_3 >
  h^b_3$, which is a contradiction since $h^b_1 = h^b_2 = h^b_3$ and
  $h^2_1 = h^2_2 = h^2_3$.
  \qed
\end{proof}

\begin{proof}[of Theorem~\ref{theorem-np}]
  For the decision version of \textsc{temporal-cut-mini\-mi\-za\-tion} we
  are given a dynamic tree $T = (T_1,\dots, T_k)$ and a number $c$ and
  must decide whether $c$ temporal cuts suffice to turn $T$ into a
  graph $T'$ with an acyclic supergraph
  $\operatorname{super}(T')$. Containment of this problem in
  $\Class{NP}$ is clear since we can simply guess the temporal cuts
  and checking whether a graph is acyclic can easily be done in
  polynomial time. 
  
  To show hardness we reduce from the $\Class{NP}$-complete problem
  $\Lang{vertex-cover}$, which contains all (coded) pairs $(G,k)$ of
  undirected graphs $G = (V,E)$ and numbers~$k$ such that there is a
  set $C \subseteq V$ with $|C| \le k$ and for all $\{u,v\} \in E$ we
  have $u \in C$ or $v \in C$. Let $G=(V,E)$ with $V =
  \{v_1,\dots,v_n\}$ and $k$ be an input for the reduction. We must
  compute an instance for \textsc{temporal-cut-minimization}
  consisting of a dynamic tree~$T$ and a number~$c$. We set $c = k$.  
  For each $v\in V$ let $v_{\text{in}}$ and $v_{\text{out}}$ be two
  new nodes and $\widehat{V}=\{v_{\text{in}},v_{\text{out}}\mid v\in
  V\}$ be the set of nodes in the supergraph of the dynamic tree. 
  For each node $v_i \in V$ let $T_i=(\widehat{V},E_i)$ with
  $E_i=\{(v_{i,\text{out}},w_{\text{in}})\mid \{v_i,w\}\in E\}$. These
  trees contain directed outgoing edges from $v_{i,\text{out}}$ to
  $w_{\text{in}}$ for each vertex~$w$ that is connected with $v_i$
  in~$G$. Finally, let $T_{|V|+1}=(\smash{\widehat{V}},E_{|V|+1})$ with 
  $(v_{\text{in}},v_{\text{out}})\in E_{n+1}$ for each node $v\in V$.
  By construction, each $T_i$ is a tree or forest (it is natural to
  allow forests, but one can also turn forests into trees by
  adding a global root and making all forest roots children of this
  global root). 
  
  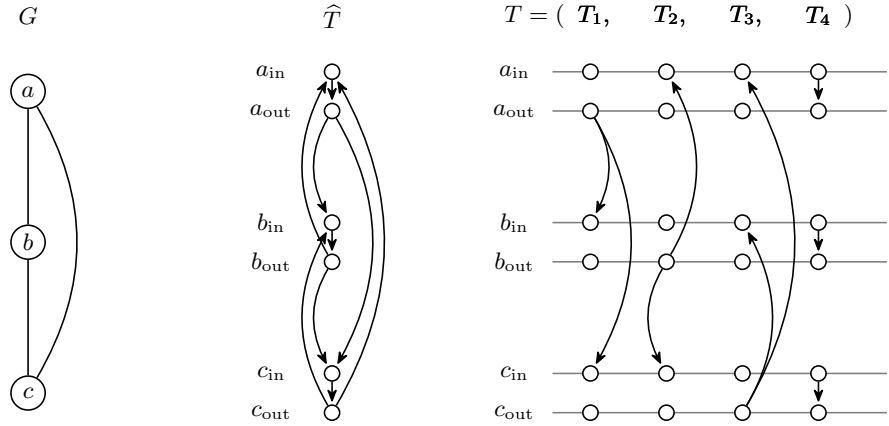
\begin{figure}[tb]
    \begin{center}
    \begin{tikzpicture}[
      graph,
      tlvertex/.style={small node},
      basearc/.style={->[bend]},
%      fbckarc/.style={-{>[bend]},red},
      fbckarc/.style={-{>}},
      timeline/.style={gray},
      ]
      % decorate,decoration={random steps,segment length=3mm,amplitude=.5pt}
      \foreach \n/\r/\y in {A/a/0cm,B/b/2cm,C/c/4cm}
      {
%        \node (\n-node)  at (-7,    -\y) [node] {$\r$};
        \node (\n-in 0)  at ( 0,-\y+.8em) {$\r_{\text{in}}$};
        \node (\n-out 0) at ( 0,-\y-.8em) {$\r_{\text{out}}$};
        \node (\n-in S)  at (-3.2,-\y+.8em) {$\r_{\text{in}}$};
        \node (\n-out S) at (-3.2,-\y-.8em) {$\r_{\text{out}}$};
        \node[node] (\n0) at (-6.4,-\y) {$\r$};
        \node[tlvertex] (\n i) at (-2.4,-\y+.8em) {};
        \node[tlvertex] (\n o) at (-2.4,-\y-.8em) {};
        \draw[timeline] (0.5cm,-\y+.8em) -- +(+4.4cm,0);
        \draw[timeline] (0.5cm,-\y-.8em) -- +(+4.4cm,0);
        \foreach \t in {1,2,3,4}
        {
          \node[tlvertex]	(\n-in\t) at (\t,-\y+.8em) {};
          \node[tlvertex]	(\n-out\t) at (\t,-\y-.8em) {};
          \node at (\t,1) {$T_\t\ifnum\t<4\rlap{,}\fi$};
        };
      }
      \node [left] at (1,1) {$T = (\ \ $};
      \node [right] at (4,1) {$\ \ )$};
      \node at (-2.4,1) {$\widehat{T}$};
      \node at (-6.4,1) {$G$};
      \draw[,,] (A0) to (B0);
      \draw[,,] (B0) to (C0);
      \draw[,bend left,] (A0) to (C0);
      
      \draw[basearc,bend right] (Ao) to (Bi);
      \draw[basearc,bend left]  (Bo) to (Ai);
      \draw[basearc,bend right] (Bo) to (Ci);
      \draw[basearc,bend left]  (Co) to (Bi);
      \draw[basearc,bend left]  (Ao) to (Ci);
      \draw[basearc,bend right] (Co) to (Ai);
      \foreach \n in {A,B,C}{
        \draw[fbckarc] (\n i) to (\n o);
        \draw[fbckarc] (\n-in4) to (\n-out4);
      };
      \draw[basearc,bend left] (A-out1) to (B-in1);
      \draw[basearc,bend left] (A-out1) to (C-in1);
      \draw[basearc,bend right] (B-out2) to (A-in2);
      \draw[basearc,bend right] (B-out2) to (C-in2);
      \draw[basearc,bend right] (C-out3) to (A-in3);
      \draw[basearc,bend right] (C-out3) to (B-in3);
    \end{tikzpicture}
  \end{center}
  \caption{Example of the reduction from $\Lang{vertex-cover}$ to
    \textsc{temporal-cut-mini\-mi\-za\-tion}. The input graph is $G =
    \bigl(\{a,b,c\}, \{\{a,b\}, \{b,c\}, \{a,c\}\}\bigr)$. The supergraph of the
    dynamic graph $T = (T_1,T_2,T_3,T_4)$ is $\widehat T$. }
  \label{figure-reduction}
  \end{figure}

  Clearly, the reduction is computable in polynomial
  time. Figure~\ref{figure-reduction} shows an example for this
  reduction.  Note that for each edge in~$G$ we get exactly one
  atomic cycle in the supergraph and all cycles in the supergraph are
  alternating paths of in-out (short, straight) and out-in (long, bend) edges.
  
  It remains to show $(G,k)\in \Lang{vertex-cover}$ if and only if
  $(T,k) \in {}$\textsc{tempo\-ral-cut-mini\-mi\-za\-tion}. Let $(G,k)$
  be an instance of $\Lang{vertex-cover}$. Then there is a vertex
  cover~$C$ of size $|C|\leq k$ for the graph $G=(V,E)$. The length of
  the sequence~$T$ is $|V|+1$.  We do the following $|C| \le k$ temporal
  cuts:
  $(v_{\text{in}},n)$ for each $v\in C$. We claim that these cuts turn
  $T$ into a new dynamic tree $T'$ for which
  $\operatorname{super}(T')$ is acyclic: By construction, no in-node
  $v_{\text{in}}$ has incoming edges in $T_{n+1}$ and all out-nodes
  have exactly one in-node. A given vertex $v_{\text{in}}$ can be
  divided by out temporal cuts into at most  two nodes
  $v_{\text{in}}$ and $v'_{\text{in}}$. This implies that if a node $v$ is
  in the vertex cover $C$, then in $T'$ neither $v_{\text{in}}$
  nor the vertex $v_{\text{out}}$ can be on a cycle. 
  If there is a cycle left, then there is an alternating path in the
  supergraph with at least two in-out-edges. Those correspond to connected
  nodes $x$ and $y$ in $G$. As the in-out-edges are on that path, neither $(x,n)$
  nor $(y,n)$ is one of our temporal cuts. Hence, neither $x$ nor $y$ are 
  in the vertex cover~$C$. Since $x$ and $y$ are connected, $C$ cannot
  be a valid vertex cover and it follows that
  $\operatorname{super}(T')$ is acyclic.  
  
  For the other direction, let there be a set $R$ or at most $k$
  temporal cuts that turn $T$ into $T'$ with
  $\operatorname{super}(T')$ being acyclic. If necessary, we replace
  all $(v_{\text{in}},j)$ and $(v_{\text{out}},j)$ in~$R$ by
  $(v_{\text{in}},n)$ since all cycles contain an in edge of
  the last tree $T_{n+1}$. Let $C=\{v\in V_G \mid (v_{\text{in}},n)\in
  R\}$. Then $C$ is a vertex cover in~$G$: If there is an edge
  $\{u,v\}$ in $G$ not covered by~$C$, then neither
  $(v_{\text{in}},n)$ nor $(u_\text{in},n)$ can be in~$R$ since,
  otherwise, the cycle
  $u_{\text{in}},u_{\text{out}},v_{\text{in}},v_{\text{out}},u_{\text{in}}$
  would still be in the supergraph. Hence, we get the claim. \qed
\end{proof}
  
\end{document}